# High Temperature Molecular Magnetism Caused by π-Electrons: Copper Phthalocyanine Doped with Alkaline Metals

E.G. SHAROYAN[*] and A.S. MANUKYAN

*Institute for Physical Researches, National Academy of Sciences, Ashtarak-2, 378410, Armenia*

**ABSTRACT:** Electron spin resonance spectra of copper phthalocyanine doped with alkaline metals ($A_xCuPc$) have been investigated. The temperature dependence of ESR spectra indicates the ferromagnetic behavior. The Curie–Weiss temperature varies from 30K to 115K depending on the stoichiometry x of samples. Some particles of polycrystalline samples were attracted to a weak magnet at temperature slightly higher than 77K. The observed magnetism is caused by unpaired π-electrons of phthalocyanine anions on the $e_g$ doubly degenerated molecular orbital. The observed ferromagnetism can be understood within the framework of the McConnell-2 model proposed for organic ferromagnetic charge-transfer complexes. The high-temperature magnetism in $A_xCuPc$ is considered to be a result of the Zener mechanism of double exchange between phthalocyanine molecular anions of different valence.

## INTRODUCTION

Among molecule-based magnets purely organic ferromagnets consisting of light elements (C, H, N, O) and not containing metallic elements are of especial interest [1-3]. The exchange interactions in purely organic magnets can be caused only by 2p-electrons. The situation is essentially different in comparison with traditional inorganic magnets, where magnetic properties are conditioned by inner d- and f-electrons. There are some difficulties connected with creation of purely organic molecular ferro- and ferrimagnetics. First, considerable difficulties arise from their chemical activity. Second, usually the antiferromagnetic ordering between neighboring molecules takes place. Third, a secondary magnetic structure in the limits of one domain is also important [4]. Nevertheless, in 1991 two communications appeared on the observation of ferromagnetism in the stable organic radical p-NPNN [5] and in the fullerene-based charge-transfer salt [TDAE]$C_{60}$ [6]. The Curie temperature in p-NPNN is 0.6K and, in spite of further many works, the values of $T_C$ in pure organic radicals are very low: they do not exceed several Kelvin's [3,7]. In case of [TDAE]$C_{60}$ $T_C$=16.1K. The increase of $T_C$ up to 17.5K at 4 kbar has been observed [8]. Many investigations were performed to reveal the nature of magnetism in this compound. Recently the Jahn–Teller distortions in the negative monoanions $C_{60}^-$ have been observed, and the cooperative Jahn–Teller effect is likely to cause ferromagnetism of this compound [2,9,10].

For the synthesis of highly conductive and magnetic materials we have developed a technology of doping of solid metal phthalocyanines (MPc, where Pc=$C_{32}H_{16}N_8$) by donors and acceptors of electrons (alkaline metals and iodine, respectively) [11,12]. A number of phthalocyanine-based charge-transfer compounds have been obtained and studied, using this technology. The alkaline-metal doped metal phthalocyanines $A_xMPc$ (at x>1.5) stand out of this group because of their unusual magnetic properties [12-17]. The most significant feature of $A_xMPc$ magnets is their high value of $T_C$ that sometimes exceeds the room temperature. Magnetic properties of $A_xMPc$ compounds vary within a broad range: they strongly depend on the extent of doping (0<x<4), the type of the central metal (M=Zn, Cu, Ni, Co,

Fe, Mn), as well as on the type of alkaline metal (A= Na, K, Rb, Cs).

In the eighties we discovered and investigated a room-temperature ferromagnetism in samples of $A_xFePc$, $A_xCoPc$, $A_xNiPc$ ($x\gg1$) [12-17]. The magnetism in these compounds can be conditioned by the mixture of MPc anions and metallic nanoparticles of Fe, Co, or Ni, which are formed during the doping of corresponding MPc. Linear dimensions of these nanoparticles are below the resolution limit of the X-ray diffraction method. It appears to be possible to change the ratio of these magnetic components, varying synthesis conditions.

In the present paper we report the ferromagnetic behavior of $A_xCuPc$ samples studied by the ESR method. The case of nanoparticle ferromagnetism in these compounds is excluded since Cu is diamagnetic metal (we do not consider here a weak Pauli paramagnetism of conducting electrons). In order to explain the unusually high values of $T_C$ in these samples, we consider the Zener mechanism of double exchange, which was previously used in explanation of the ferromagnetism in inorganic crystals with mixed-valence ions [18,19]. Taking into account the growing interest towards magnetic properties of alkaline doped phthalocyanines we also present different methods of their heterophase doping.

**DOPING METHODS. HETEROPHASE SYNTHESIS OF SAMPLES**

The doping of MPc with alkaline metals leads to the reduction of MPc molecules [20–25]. In the solution (usually tetrahydrofuran (THF) is used) a step-by-step reduction of mono-, di-, tri- and tera-anions of MPc takes place that can be presented by the following scheme:

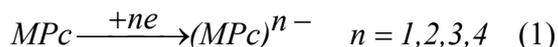
$$MPc \xrightarrow{+ne} (MPc)^{n-} \quad n = 1,2,3,4 \quad (1)$$

The doping of MPc in a solid phase can be presented by the analogous reaction:

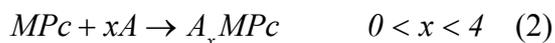
$$MPc + xA \rightarrow A_xMPc \quad 0 < x < 4 \quad (2)$$

However, compounds obtained by the doping of monocrystals, polycrystals or films of MPc significantly differ from the products of the liquid-phase synthesis. With doping of MPc in the solution one can obtain the $(MPc)^{n-}$ anions only with the integer charge n=1,2,3,4. With doping of solid phthalocyanines one can obtain molecules also with fractional charges (for instance, at n=1.5, 2.5, and so on). Here an opportunity appears to obtain molecules (anions) with different valence (see below), as well as phases with different stoichiometries; in addition, a "layering" of phases is possible analogous to solid solutions of manganites [26,27]. Varying the value of x, one can obtain a number of compounds with different structural and, correspondingly, electrical and magnetic properties.

For the synthesis of $A_xMPc$ compounds in the wide interval of x (0<x<4), we used the intercalation technique of doping. The scheme of the apparatus for the synthesis and the temperature profiles along the reaction ampoule are shown in Fig.1. Here $T_A$ and $T_{MPc}$ are the temperatures of the alkaline metal and MPc, respectively. Three different regimes were proposed and developed in our laboratory: regime 1, when $T_A > T_{MPc}$, regime 2 with $T_A=T_{MPc}$, and regime 3, when $T_A<T_{MPc}$. The first regime leads to the condensation and formation of the liquid phase of the alkaline metal on the surface of phthalocyanines. It corresponds to the doping from the liquid phase (so-called liquid-phase doping). In the third regime the condensation and formation of the liquid phase are impossible and it corresponds to the doping from the gas phase (so-called gas-phase doping). For obtaining $A_xMPc$ samples in the present work we used the regime 2. In this case the results of doping strongly depend also on the geometry of the reaction ampoule, weights of A and MPc, etc. Polycrystalline samples of copper phthalocyanines (α-CuPc and β-CuPc) were doped with sodium (Na) and rubidium (Rb). More than 30 samples have been synthesized. The synthesis temperature for $Na_xCuPc$ was $350^0C$ and for $Rb_xCuPc$ it was $220^0C$. The reaction time was varied between 2 and 7 hours for $Na_xCuPc$ and between 15 minutes and several hours for $Rb_xCuPc$. It seems to be difficult to obtain uniformly doped samples, if it is possible in principle. For obtaining "uniformly" doped samples and removing the excess amounts of alkaline metal (deintercalation) different regimes of annealing have been used.

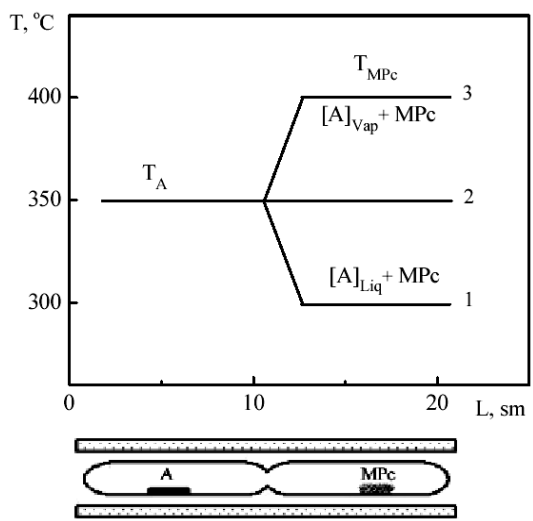

**Fig.1.** Scheme of the apparatus for synthesis and temperature profiles along the reaction ampoule.

**TEMPERATURE DEPENDENCE OF ESR SPECTRA**

For magnetic investigations we used a conventional X-range ESR-spectrometer, which is the most informative in determination of the degree of reduction of MPc. The ratio of integrated ESR-signal intensities at T=77K and T=293K was measured for each sample. Several samples which exhibited the strong deviation from the Curie law have been studied in a wide temperature interval from 3.5K to 300K using a "JEOL-JES-PE-3X" radio-spectrometer. The main results of ESR measurements are presented in Figs.2,3. Figure 2a shows the ESR spectra of polycrystalline samples of α-CuPc and β-CuPc. These spectra well coincide with the known literature data. The ESR signal is caused by the $Cu^{2+}(d^9)$ ions on the $b_{1g}(d_x^2-d_y^2)$ molecular orbital (MO) as it is shown in Fig.4. During doping the intensity of this "wide" signal decreases monotonously and a narrow signal at g~2 arises, the intensity of which raises (Fig.2b). The final integral intensity of the narrow signal is about $10^{20}$spin/g. This signal is associated with π-electrons on the degenerate $e_g$ MO (Fig.4). Temperature dependences of the magnetic susceptibility ($I_{ESR}T$ and $1/I_{ESR}$) of $Na_xCuPc$ samples are presented in Fig.3.

The following features are worth mentioning.
1) In samples 2 and 3 the temperature dependence of the integral intensity of ESR spectra deviates essentially from Curie law. In Fig.3a the dotted horizontal line corresponds to Curie law ($I_{ESR}T$=const=1). The constant is normalized to $I_{ESR}(293)\cdot 293$. These essential deviations from Curie law indicate strong ferromagnetic interactions. Similar results were also obtained for $Rb_xCuPc$ samples.
2) The mixture of different stoichiometric phases within one sample leads to the complex temperature dependence of $I_{ESR}T$ and $1/I_{ESR}$. Two values of the Curie–Weiss constant were calculated for sample 3: $\theta_1$=115K and $\theta_2$=40K. Some particles of this sample were attracted to the small magnet at the temperature slightly higher than 77K that also confirms that $T_C$ of these particles is above 77K. For sample 2 $\theta_3$=50K. The values of $\theta_1, \theta_2$ and $\theta_3$ exceed significantly $T_C$ of existing purely organic ferromagnetic compounds. Different $\theta$ values observed from the same polycrystalline sample can be explained by the nonuniform doping. Indeed, elemental analysis done by INGA - Energy 300 (Electron Dispersive Spectrometer) indicated that x values vary from 1 to 4 in different parts of the same sample. The average value of x in the samples 2 and 3 is about 2.5.
3) The linear decrease in the $I_{ESR}T$ product was observed at low temperatures. Such behavior may be explained by the saturation of magnetization at temperatures below 25–30K in case of the ferromagnetic ordering. The similar behavior of the $I_{ESR}T$ product may also be expected in case of a ferromagnetic cluster glass. Additional studies of the temperature dependence of the AC magnetic susceptibility and magnetization are needed for the final analysis.

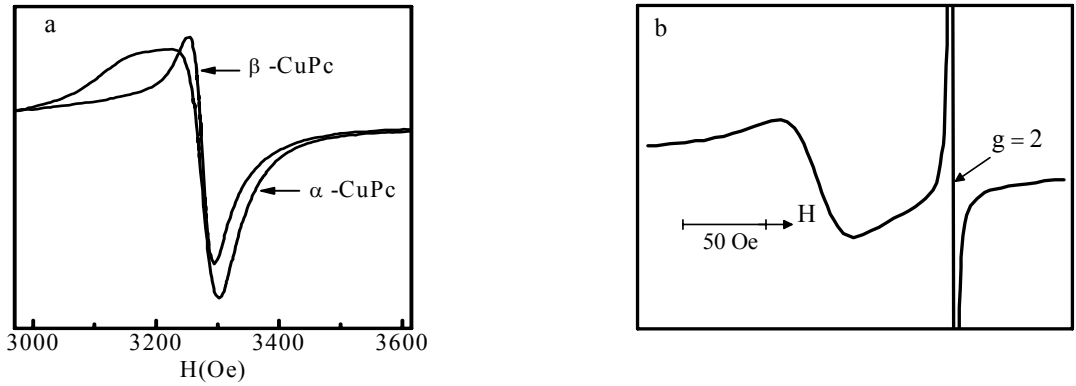

Fig.2. a) ESR spectra of α-CuPc and β-CuPc,   b) ESR spectra of Na$_x$CuPc (x<2) samples.

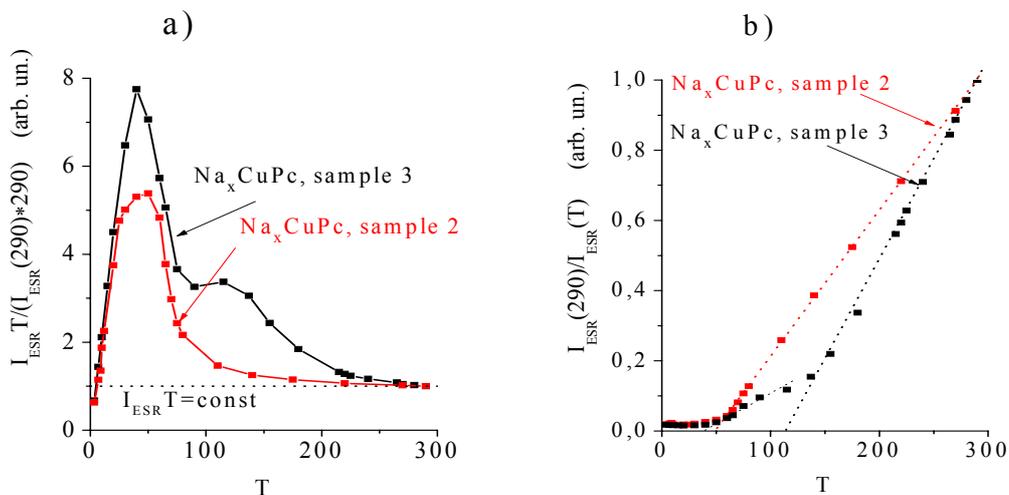

Fig.3. a) Dependence of ($I_{ESR}T$) on $T$,    b) Dependence of $1/I_{ESR}$ on $T$.

## MIXED VALENCE. DOUBLE EXCHANGE. COMBINED EFFECT OF MAGNETIC AND DOUBLE EXCHANGES

The present work and our previous studies show that the temperature dependence of the integral intensity of the ESR signals in A$_x$CuPc deviates essentially from the Curie law. The maximal deviations occur at x =2.5. In order to explain magnetic properties of alkali-doped CuPc, we consider the energy levels of CuPc - Fig.4 [28,29]. The case x=2 corresponds to the [Cu$^I$Pc (-3)]$^{2-}$ dianion in which d-orbitals are completely filled and one π-electron occupies the double degenerated e$_g$ molecular orbital. The case x=2.5 may correspond to molecules having different valence, with one and two π-electrons, respectively. Figure 4 presents the scheme of molecular orbitals of two adjacent anions with different valence at x=2.5. In the ground state their total spin S=3/2, because the virtual transitions with the charge transfer stabilize the ferromagnetic coupling. This is, in fact, the Mc-Connel-2 model (Wudl's modification) for organic ferromagnetics with a metallic conductivity [30,31].

The system π$^1$–π$^2$ strongly resembles inorganic magnetic dimmers (d$^1$–d$^2$) in which the mechanism of double exchange is essential [32]. The delocalization of an "extra" electron between adjacent molecular anions with the spin S$_0$=1/2 leads to their ferromagnetic ordering with S=3/2. It should be noted that the stabilization of the ferromagnetic state due to the double exchange is much stronger than the Heisenberg stabilization [32].

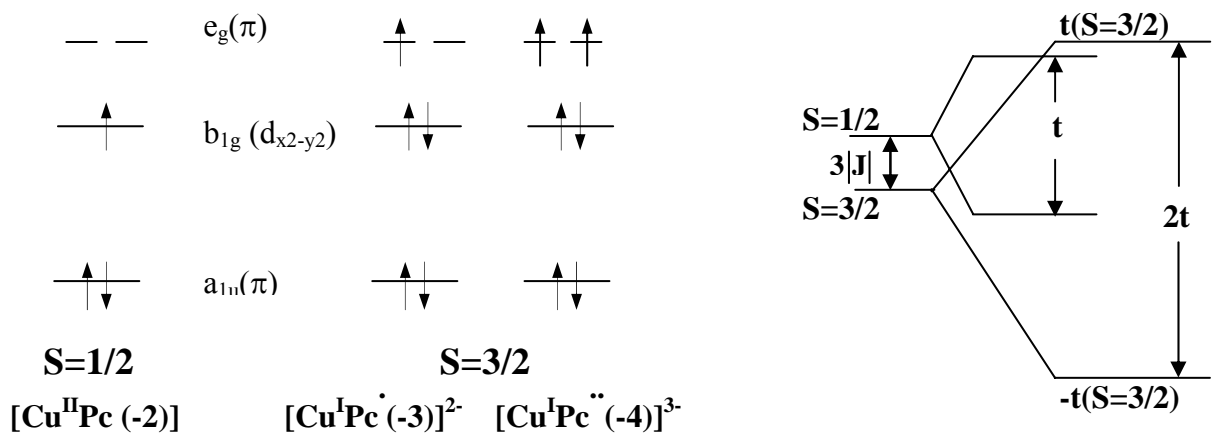

Fig.4. Scheme of valence molecular orbitals of CuPc and adjacent anions $[Cu^IPc^{\cdot}(-3)]^{2-}$ and $[Cu^IPc^{\cdot\cdot}(-4)]^{3-}$ at x=2.5.

Fig.5. Combined action of the magnetic and double exchanges at t/J=6.

The combined action of the magnetic exchange and double exchange results in the following energy levels:

$$E = -JS(S+1) + t(S+1/2)(2S_0+1), \quad (3)$$

where J is the many-electron exchange parameter, t is the transfer integral, and $S_0$ is the value of the spin core [32]. In case of the $\pi^1-\pi^2$ system both J and t are greater than zero. Both interactions make a ferromagnetic contribution. The result of combined effect of the magnetic exchange and double exchange is shown in Fig.5. Here the case is presented, when the ratio t/J=6. For estimations we took t=0.3 eV, J=0.05 eV ($U_{eff}$=1.5 eV). Thus, in the $A_xMPc$ compounds the double exchange, strongly stabilizing the ferromagnetic state, leads to high values of $T_C$.

## CONCLUSION

At doping of the copper phthalocyanine with alkaline metals we have obtained ferromagnetic compounds, where the Curie temperature is higher than 77K. In $A_{2.5}CuPc$ the d-orbitals of the Cu(I) ions are completely filled ($d^{10}$). We believe that the high-temperature ferromagnetic behavior of $A_xCuPc$ samples is conditioned by π-electrons on the double degenerated $e_g$ molecular orbital, mixed valence of CuPc anions and mechanism of double exchange, which is first observed in molecular ferromagnetics. The values of $T_C$ in $A_xCuPc$ significantly exceed the Curie temperatures of all known purely organic compounds.

**Acknowledgments**

The authors are grateful to H.R.Asatryan for the assistance in low-temperature ERS measurements at A.Ioffe Physical-Technical Institute (St. Petersburg), as well as to A.H.Maloyan and S.G.Gevorgyan for useful discussions. This work was supported by the ANSEF grant no.PS139–01.